# Enhancing ICT Literacy and Sustainable Practices in the Hospitality Industry: Insights from Mnquma Municipality

Jose Lukose and Abayomi Agbeyangi

Walter Sisulu University, Buffalo City Campus, East London, South Africa
Corresponding: aagbeyangi@wsu.ac.za

**Abstract.** The leisure and hospitality industry is a significant driver of the global economy, with the adoption of new technologies transforming service delivery and customer experience. Despite the transformative potential and benefits associated with adopting technology, there remains a low level of adoption in rural areas, particularly among small-scale players. This study explores the role of ICT literacy and sustainable practices in influencing ICT adoption among small-scale players in the hospitality industry in rural Eastern Cape Province, South Africa, specifically focusing on Mnquma Municipality. The study employs a non-probability sampling and purposive technique, utilising a case study research design within a positivist paradigm. A random sample of 21 small-scale players (BnBs, guest houses, and non-serviced accommodations) was selected, and data were collected through a face-to-face interview and questionnaire featuring closed-ended questions. The data were analysed using descriptive statistics and the Kruskal-Wallis H Test to examine differences in ICT usage levels. The test yielded a Kruskal-Wallis H of 2.57 with a p-value of 0.277. The findings reveal that businesses with more educated workforces demonstrate higher ICT adoption levels. Moreover, key factors such as ICT literacy, awareness of sustainable practices, access to ICT resources, and contextual challenges significantly impact ICT adoption. Recommendations include integrating ICT literacy and sustainability education into training programs and developing targeted policies and support mechanisms to enhance ICT integration.

## 1. Introduction

The hospitality industry is a cornerstone of economies worldwide, encompassing a broad spectrum of services such as hotels, restaurants, event planning, and transportation. Within this industry, small and medium-sized hospitality organisations (SMHOs) play a pivotal role, often family-owned and deeply rooted in their local communities (Ofori-Amanfo et al., 2022). These enterprises contribute significantly to economic activity and serve as crucial providers of employment and cultural enrichment (Dogru et al., 2020; Thommandru et al., 2023). Delivering high-quality service is central to the success of hospitality businesses, which is essential for building customer loyalty and sustaining profitability (Karim & Rabiul, 2024; Matsuoka, 2022). In an era of rapid technological advancements, information and communication technologies (ICTs) have emerged as critical tools for enhancing operational efficiency and strategic decision-making within the hospitality sector (Anser et al., 2020; Melián-Alzola et al., 2020). Melián-Alzola et al. (2020) underscored the significance of information technology investments as pivotal in augmenting the agility of hotels to adapt to environmental fluctuations. Consequently, adopting digital transformation practices has emerged as an imperative for hotels aiming to maintain their competitive edge within the industry.

The integration of ICTs offers SMHOs numerous benefits across various facets of management, including operations management, financial modelling, growth management, and customer relationship management (Lunkes et al., 2020; Siti-Nabiha et al., 2021). These technologies streamline internal processes, enable businesses to adapt to changing market demands and enhance their competitive edge in the industry. However, successful ICT adoption is not solely dependent on the availability of technology but also on the level of ICT literacy among employees and business owners. Providing adequate training and educational resources is crucial for ensuring staff members can effectively use new technologies, maximising their potential benefits (Dayour et al., 2023).

Despite its transformative potential, the adoption of ICTs by small-scale players in the hospitality industry within specific regions, such as the Mnquma Municipality in the Rural Eastern Cape Province of South Africa, remains underexplored. Many small-scale hospitality enterprises in these regions are family-operated and may face barriers such as limited technology access and inadequate ICT utilisation training. Understanding these challenges and exploring the factors influencing ICT adoption is crucial for developing strategies that assist these businesses in effectively leveraging technology to improve their operational capabilities and sustainability. This study seeks to investigate the specific factors influencing the adoption of ICTs among small-scale players. It aims to assess the level of awareness among these operators regarding the benefits of ICT literacy and adoption in their businesses. This is intended to uncover the unique challenges and opportunities associated with ICT literacy and adoption to ensure sustainability in this context. The following research questions serve as a guide:

- What role does education play in facilitating ICT adoption and sustainable practices within the hospitality industry?
- What factors influence ICT adoption among small-scale hospitality players?
- What barriers exist to accessing and utilising ICT resources?

The Mnquma Municipality[1], situated within the Amathole district of the Eastern Cape Province, South Africa, is a predominantly rural region. The area incorporates several towns, including Butterworth, Centane, and Ngqamakhwe. Despite its natural beauty and attractions, such as the Bawa Falls and Blythswood rock art, Mnquma faces challenges that are common to many rural areas, including limited access to resources and infrastructure. This context makes it a particularly interesting case study for investigating the adoption of information

---

[1] https://www.mnquma.gov.za/

and communication technology (ICT) within the local, small-scale hospitality industry.

## 2. Literature Review

In the hospitality industry, ICT plays a crucial role in enhancing service delivery (Melián-Alzola et al., 2020), operational efficiency (Mercan et al., 2021), and customer satisfaction (Matsuoka, 2022). This capability is vital for small and medium-sized hospitality organisations (SMHOs), which rely heavily on personalised service to maintain a competitive edge. The implementation of ICT enables these businesses to streamline operations, improve customer interactions, and ultimately enhance service quality. According to Mercan et al. (2021), in the hospitality sector, ICT can facilitate personalised services, enhanced customer experiences, and innovative product offerings, leading to increased revenue and decreased costs. The increasing adoption has been acknowledged as a significant advancement in recent years, providing advantages such as enhanced service delivery, operational efficiency, and cost savings (Ip et al., 2011). Nevertheless, the rate of ICT adoption varies, and several studies have indicated low adoption rates in specific regions (Mwai, 2016; Nwakanma et al., 2014). Despite challenges, ICT is considered an important aspect of the hospitality industry, contributing to business success by widening the client base and facilitating customer feedback (Richard, 2012). Moreover, ICT adoption is increasingly linked to enhancing educational resources within the hospitality sector, promoting technical efficiency, knowledge-sharing, and training opportunities (Dayour et al., 2023).

### 2.1 Factors Influencing ICT Adoption

The integration of ICT in the hospitality industry is influenced by various factors that can either enable or impede its adoption. One of the pivotal factors is the availability of resources. Small and medium-sized hospitality organisations (SMHOs) often operate with limited financial and technological resources, hindering their capacity to invest in advanced ICT solutions. As stated by Cheng et al. (2021), access to financial resources significantly influences the adoption of ICT, as it enables businesses to procure the necessary hardware, software, and infrastructure. Without sufficient financial support, numerous SMHOs encounter challenges in keeping pace with technological advancements, leading to slower adoption rates and potential competitive disadvantages. In addition to financial constraints, the availability of educational resources and ICT training programs significantly affects adoption rates. SMHOs that invest in continuous training and development for their employees tend to exhibit higher adoption rates, as these initiatives foster a workforce capable of leveraging new technologies effectively (Wang et al., 2020).

More so, education and training of the workforce play a significant role in the adoption of ICT. The level of education and ICT literacy among employees directly impacts the effective utilisation of these technologies. Businesses with a well-educated workforce and access to ongoing training programs are more inclined to adopt and successfully implement ICT. According to Wang et al. (2020), regular training sessions and workshops are crucial in enhancing employees' ICT skills and knowledge. This continuous learning approach ensures that staff excel in using current technologies and remain prepared to embrace new and emerging ICT tools. As highlighted by Kyakulumbye and Pather (2022), ICT adoption in hospitality is facilitated by fostering a culture of learning and innovation within organisations, which can be achieved through partnerships with educational institutions and vocational training centres.

Similarly, management support and commitment are essential for successfully adopting ICT or any innovation (Barham et al., 2020). The perspective and leadership of business owners and managers towards technology significantly impact the integration of ICT into business operations. According to Melián-Alzola et al. (2020), strong leadership and a well-defined strategic vision for ICT adoption can facilitate the successful implementation of these technologies. Managers who recognise the strategic advantages of ICT are more likely to allocate resources, inspire employees, and foster a culture that embraces technological innovation. In addition, seeking external support and forming partnerships with technology providers, industry associations, and government agencies can help facilitate the adoption of ICT (Kyakulumbye & Pather, 2022). External support allows businesses to tap into expertise, funding, and resources that may not be available internally. Government initiatives and subsidies aimed at promoting ICT adoption can also significantly reduce barriers for small- and medium-sized hospitality organisations (Barham et al., 2020). Collaborating with technology providers can result in tailored solutions that meet the specific needs of small-scale businesses in the hospitality industry. These partnerships can provide continuous technical support, ensuring that businesses can effectively maintain and enhance their ICT systems.

Another aspect is that perceived benefits and competitive pressure significantly motivate businesses to adopt ICT (Karim & Rabiul, 2024). Businesses that recognise clear advantages in adopting ICT, such as increased efficiency, improved customer service, and higher revenue, are more inclined to invest in these technologies. Additionally, competitive pressure within the industry drives businesses to keep abreast of technological advancements to maintain their market position. According to (Matsuoka, 2022), the imperative to remain competitive in a technology-driven market compels many businesses to adopt ICT to differentiate themselves and provide superior services. Hence, the adoption of ICT in the hospitality sector is complex and shaped by factors such as resource availability, employee education and training, managerial backing, external partnerships, and perceived advantages. Small and medium-sized hospitality operators must thoroughly address these factors to seamlessly integrate ICT, ensuring they possess the required resources, expertise, and support to utilise technology effectively. This approach will enable these businesses to boost operational efficiency, enhance service delivery, and stay competitive in an ever-evolving digital market. For example, acquiring expensive software licenses and advanced technological hardware can financially burden small-scale hospitality businesses in rural areas. Additionally, the lack of ICT skills among employees, who are often hired based on their ability to deliver traditional hospitality services rather than technological proficiency, can further hamper adoption efforts (Kyakulumbye & Pather, 2022). Addressing these barriers requires targeted interventions, such as affordable technology solutions, training programmes, and improved access to digital infrastructure. Addressing these barriers requires targeted interventions, such as affordable technology solutions, training programs, and improved access to digital infrastructure. Furthermore, educational initiatives tailored specifically to the needs of rural SMHOs can significantly improve ICT literacy and adoption rates.

## 2.2 ICT Adoption Models and Theories

Researchers have developed several models and theories to understand and predict ICT adoption. The Diffusion of Innovation Theory (IDT) (Miller, 2015) explains the adoption process as an S-shaped curve influenced by economic, psychological, and sociological factors. Based on the readiness to adopt new technologies, this theory categorises adopters into innovators, early adopters, early majority, late majority, and laggards (Dillon & Morris, 1996). Another prominent model is the Technology Acceptance Model (TAM) (Davis, 1989), which focuses on individual perceptions and attitudes towards new technology derived from the Theory of Reasoned Action. TAM posits that perceived ease of use and perceived usefulness are key determinants of technology adoption. The Unified Theory of Acceptance and Use of Technology (UTAUT), as proposed by (Venkatesh et al., 2003), builds on the Technology Acceptance Model (TAM) by integrating supplementary elements, including social influence, facilitating conditions, and user experience. These models provide valuable frameworks for understanding the dynamics of ICT adoption in various contexts, including the hospitality industry. In the context of SMHOs, these models also underscore the importance of education and training in shaping perceptions of usefulness and ease of use. Enhancing employee ICT literacy through continuous education can bridge the gap between technology's perceived complexity and its actual application (Wang et al., 2020).

The diffusion of innovation theory encompasses various factors that influence the adoption of innovations in a broader context. These factors include relative advantage, compatibility, complexity, trialability, and observability. Relative advantage refers to the perceived benefits of the innovation compared to existing solutions. As noted by Goh and Sigala (2020) in their study on integrating ICT into classroom instruction, the perceived relative advantage refers to when a lecturer perceives that innovation offers greater benefits than existing methods. For example, if lecturers find new technology more convenient and visually beneficial, like using Microsoft PowerPoint over projector transparency slides, they are more likely to adopt it. Similarly, compatibility is the degree to which the innovation aligns with existing values, past experiences, and the needs of potential adopters. Complexity involves the perceived difficulty of understanding and using the innovation. Trialability is the extent to which the innovation can be experimented with on a limited basis, and observability refers to the visibility of the innovation's results. In the hospitality industry, adopting ICT solutions is more likely if they demonstrate clear advantages, align with business operations, are easy to use, can be trialled, and show visible benefits (Goh & Sigala, 2020; Hsu, 2016).

The Technology Acceptance Model (TAM) identifies two primary factors influencing the adoption of new technology: perceived usefulness and ease of use (Davis, 1989). Perceived usefulness refers to how much a person believes that using a specific system will improve their job performance, while perceived ease of use refers to their belief that using the system will be effortless. In the hospitality industry, if business owners and employees view ICT as beneficial for enhancing service delivery and operational efficiency and find it user-friendly, they are more likely to embrace it. Rafique et al. (2020) state that TAM helps understand factors influencing technology acceptance and provides a framework to assess perceived usefulness and ease of use. Liu et al. (2022) affirm that TAM helps understand individuals' intentions and behaviour towards technology usage and can be applied in various fields like tourism to predict and explain behavioural intention. Educational interventions to increase perceived ease of use can play a critical role in ICT adoption by reducing the perceived barriers associated with complex technologies.

The Unified Theory of Acceptance and Use of Technology (UTAUT) builds on TAM by incorporating additional factors such as social influence, facilitating conditions, and experience (Venkatesh et al., 2003). Social influence refers to the degree to which individuals perceive that others who are important to them believe they should use technology. Facilitating conditions refer to the perceived availability of resources and support for using the technology. For example, facilitating conditions such as the availability of ongoing ICT training, mentorship programs, and access to user-friendly tools can significantly improve adoption rates among SMHOs (Dayour et al., 2023). These factors can collectively influence behavioural intention and user behaviour. In the hospitality industry, if employees perceive that their peers and supervisors encourage ICT use and if they have access to the necessary resources and support, their intention to use ICT will be positively affected. Research has demonstrated that UTAUT is a comprehensive model that provides insights into understanding users' adoption behaviour, especially in complex environments where multiple factors influence decisions (Rafique et al., 2020).

Research on ICT adoption in small-scale businesses highlights both the benefits and challenges of technology integration. Studies have shown that ICT can significantly enhance operational efficiency (Mercan et al., 2021), customer service (Matsuoka, 2022), and overall business performance (Melián-Alzola et al., 2020). In the hospitality industry, ICT enables better market responsiveness and operational improvements, such as efficient booking systems, customer relationship management, and streamlined supply chain operations (Anser et al., 2020; Mercan et al., 2021). However, small-scale businesses, particularly rural ones, often struggle with the costs and complexities of implementing and maintaining ICT systems (Sahadev & Islam, 2005).

Despite extensive research on ICT adoption, significant gaps remain, particularly concerning small-scale hospitality businesses in rural areas of developing countries. Most existing studies have predominantly focused on urban settings or larger enterprises, leaving a critical void in understanding the distinct challenges and opportunities rural SMEs face. One particularly underexplored area is the role of education and training in facilitating ICT adoption. For small-scale businesses in rural settings, the lack of ICT literacy and technical expertise among staff can be a significant barrier to effectively integrating technology into their operations. Research on the effectiveness of ICT adoption strategies tailored to these unique needs—including educational interventions that enhance ICT literacy and practical skills—is notably scarce. Addressing these gaps requires targeted studies that consider rural areas' socio-economic and infrastructural realities and the critical need for educational support.

## 3. Methods

This study employs non-probability sampling and purposive sampling techniques, utilising a case study research design within a positivist paradigm. The focus is on understanding the factors influencing ICT adoption among small-scale hospitality businesses in the Mnquma Municipality, Eastern Cape Province, South Africa. This research is particularly significant as it addresses the educational aspects of ICT adoption, seeking to understand how training and knowledge dissemination impact technology integration in these rural hospitality businesses.

### 3.1 Data Collection

The sample comprises 21 small-scale businesses within the rural community, including BnBs, guest houses, and non-serviced accommodations (self-catering). Data were collected using a structured questionnaire featuring closed-ended questions designed to capture information on ICT adoption, awareness, barriers, perceived benefits, and the educational background of the respondents. The questionnaire was administered through face-to-face interviews and surveys to ensure comprehensive data collection. Special attention was given to understanding the role of ICT literacy and training in these businesses. A non-parametric test (Kruskal-Wallis H test) was used alongside descriptive statistics to analyse the data. Non-parametric tests are suitable as they do not assume a normal distribution and are robust for the sample. Participants were informed about the study's purpose and provided informed consent, with assurances of confidentiality and the right to withdraw at any time. The study also incorporated questions on previous exposure to ICT training programs and the availability of ongoing training opportunities. All data was anonymised and securely stored, ensuring compliance with ethical standards.

### 3.2 Formalisation

The Kruskal-Wallis H test is used to determine whether there are statistically significant differences in the distribution of the dependent variable across the three independent groups. This test was particularly chosen to examine differences in ICT adoption levels based on varying degrees of ICT literacy and educational background among the respondents.

The hypotheses for the Kruskal-Wallis H Test are as follows:

- Null Hypothesis ($H_0$): The distribution of the dependent variable is the same across the different groups.
- Alternative Hypothesis ($H_1$): At least one group's distribution differs from the others.

The test statistic $H$ for the Kruskal-Wallis H Test is calculated using the following formula:

$$H = \left(\frac{12}{N(N+1)} \sum \frac{R_i^2}{n_i}\right) - 3(N+1)$$

Where:
- $N$ is the total number of observations across all groups.
- $R_i^2$ is the sum of ranks for the $i$-th group.
- $n_i$ is the number of observations in the $i$-th group.

The test statistic $H$ follows a chi-square distribution with $k-1$ degrees of freedom, where $k$ is the number of groups.

## 4. Results and Discussion

The findings from the data analysis focus on the level of ICT literacy and awareness, access to ICT resources, and factors influencing ICT adoption among small-scale players in the hospitality industry.

### 4.1 Demographic Information

The demographic analysis (see Table 1) of the respondents reveals a diverse age range, with the majority falling between 25 and 29 years (38%), followed by those aged 30 and 40 years (33%), and a smaller proportion aged 20–24 years (19%) and 40–49 years (10%). Gender distribution indicates a predominance of female respondents, who constitute 81% of the sample. In comparison, males account for 19%, suggesting that the small-scale hospitality businesses in the area are largely managed or owned by women. The types of businesses include bed and breakfasts (BnBs), guest houses, and non-serviced accommodations (self-catering), encompassing various classifications within the hospitality industry. The employment levels varied, covering both management and operational staff, providing a diverse view of the workforce dynamics.

Table 1: Demographic Characteristics of Respondents

| Level of Awareness | | Percentage (%) |
|---|---|---|
| Age | 20-24 | 19 |
| | 25-29 | 38 |
| | 30-40 | 33 |
| | 40-49 | 10 |
| | >50 | 0 |
| Gender | Male | 81 |
| | Female | 19 |
| Business Type | BnB | 38 |
| | Guest House | 43 |
| | Non-Serviced | 19 |
| Employment Level | General Worker | 14 |
| | Manager | 29 |
| | Owner | 57 |

## 4.2 Level of Awareness of ICT Benefits

The respondents' awareness of ICT benefits was assessed through various indicators such as years of ICT usage, impact on front office operations, and utilisation of ICT tools. Table 2 shows that most respondents have used ICT for at least a year, with nearly half having 1-2 years of experience. This highlights a general familiarity with and positive reception of ICT. All respondents unanimously reported positive effects of ICT on their business operations, suggesting a high level of awareness regarding its benefits. The use of ICT in business is prevalent, with 95% of respondents incorporating it into their operations. Social media platforms, especially Facebook, are the most commonly used ICT tools for marketing purposes, followed by personal websites. A significant portion of the businesses (67%) are aware of e-marketplaces, and 48% actively participate in them, showing a proactive approach towards leveraging digital platforms.

Table 2: Level of Awareness of ICT

| Level of Awareness | | Percentage (%) |
|---|---|---|
| Years of Usage | <1 year | 19 |
| | 1 – 2 years | 48 |
| | 3 – 5 years | 28 |
| | >5 years | 5 |
| ICT in Business | Use | 95 |
| | Does not use | 5 |
| ICT Tools in Business | Gumtree (advertising) | 10 |
| | WhatsApp (customer interaction) | 5 |
| | Facebook (marketing) | 57 |
| | Email (communication) | 5 |
| | Own website | 19 |
| Participation in E-markets | Aware of e-marketplaces | 67 |
| | Participates in e-marketplaces | 48 |

## 4.3 Level of Access to ICT Resources

Access to ICT resources was evaluated based on the availability of devices, internet access, and information storage methods. It appears widespread among the respondents, as shown in Table 3. Most have access to personal computers, laptops, and smartphones. Internet access is almost universal, with 95% of respondents connected. The primary devices for accessing the internet are personal computers and laptops. Wi-Fi availability for customers is common, with various models of internet access ranging from pay-as-you-go (24%) to free in-room Wi-Fi. Information storage and transaction recording are primarily done electronically (43%) or through a combination of electronic and traditional pen-and-paper methods (52%). This hybrid approach suggests a transition phase where businesses gradually move towards full digitalisation.

Table 3: Level of Access to ICT Resources

| Access to ICT Resources | | Percentage (%) |
|---|---|---|
| Device | Personal Computer | 19 |
| | Laptop | 33 |
| | Tablet | 5 |
| | Smartphone | 19 |
| | Business Phone | 19.0% |
| | None | 5 |
| Internet Access | Yes | 95 |
| | No | 5 |
| Wi-Fi Option | Pay-As-You-Go | 24 |
| | Free In-Room | 14 |
| | Through Front Desk | 14 |
| | Others | 43 |
| Information Storage | Electronic | 43 |
| | Manual | 5 |
| | Both | 52 |

## 4.4 Factors Influencing ICT Adoption

Several factors influencing ICT adoption were examined, including internet speed and affordability, workforce educational level, computer literacy, ICT skills, and the availability of ICT training. The findings emphasise the crucial role of education in ICT adoption. As shown in Table 4, the findings indicated that while internet access was widespread, speed and affordability varied among respondents. Internet connectivity is generally reliable, with 86% of respondents having access. However, the speed and affordability of internet services vary. While 48% find the internet affordable, 38% consider it expensive, indicating cost as a potential barrier to ICT adoption.

A key educational aspect is the workforce's educational level, predominantly high school (62%), with a good number having some form of tertiary education (33%). This higher educational attainment directly correlates with high computer literacy among respondents, with 86% reporting computer-literate employees. Additionally, most respondents rate their ICT skills as moderate (52%) to good (24%), reflecting the benefits of their educational background. Nearly half of the respondents (48%) have received formal training in ICT usage, often organised by the business owners or pursued out of personal interest.

ICT decisions are primarily made by the owner or an ICT department, and ICT maintenance is mostly outsourced to external companies. All three barriers (limited understanding of ICT usage, inadequate ICT resources, and lack of capital for ICT infrastructure) are encountered at a similar rate (33%). This implies that addressing educational gaps and providing targeted ICT training could significantly enhance ICT adoption rates among small-scale players in the hospitality industry.

Table 4: Factors Influencing ICT Adoption

| Factors on Adoption | | Percentage (%) |
|---|---|---|
| Reliable Internet | Yes | 86 |
| | No | 14 |
| Internet Speed | Very Fast | 19 |
| | Fast | 24 |
| | Manageable | 47 |
| | Very Slow | 5 |
| Internet Affordability | Very Expensive | 12 |
| | Expensive | 38 |
| | Affordable | 48 |
| Workforce Educational Level | Elementary | 5 |
| | High School | 62 |
| | Tertiary | 33 |
| Computer Literacy | Literate | 86 |
| | Non-literate | 14 |
| ICT Skills Rating | Very Good | 24 |
| | Good | 24 |
| | Moderate | 52 |
| Training on ICT Usage | Received | 48 |
| | Not Received | 52 |
| ICT Barrier | Limited Understanding | 33 |
| | Inadequate Resources | 33 |
| | Lack Of Capital | 33 |

### 4.5 The Kruskal-Wallis H Test

The Kruskal-Wallis H test was applied to examine whether significant differences exist across the two categories (years of usage and types of businesses). The test yielded a Kruskal-Wallis H statistic of 2.57 with a p-value of 0.277. The p-value greater than 0.05 indicates that the null hypothesis is rejected, suggesting there is no statistically significant difference in the level of ICT usage among the different categories of years of usage. This implies that the duration of ICT usage (whether less than a year, 1-2 years, 3-5 years, or more than 5 years) does not significantly impact the level of ICT usage among the respondents. This suggests that factors other than duration, such as educational attainment, accessibility to ICT resources, availability of training, and individual business needs, might play a more significant role in influencing the adoption and use of ICT tools. The uniformity in ICT usage across different experience levels underscores the pervasive nature of ICT benefits and its essential role in modern business operations. Figure 1 illustrates the distribution of the factors influencing the levels of adoption by year of usage.

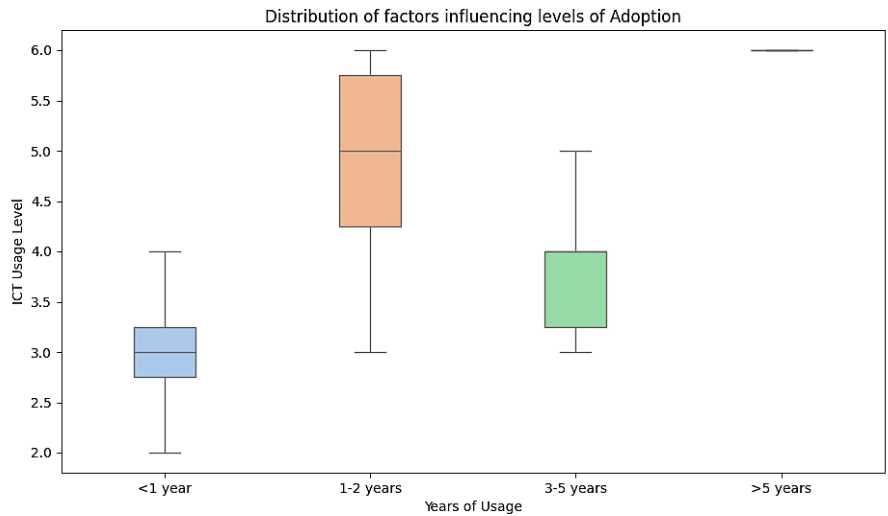

**Figure 1: Distribution of ICT Usage Levels by Years of Usage**

Figure 2 illustrates the distribution of ICT usage levels across the three different types of businesses: BnBs, guest houses, and non-serviced accommodations (self-catering). The box plot displays each group's median, quartiles, and potential outliers. BnBs show a relatively narrow interquartile range (IQR), indicating consistent ICT usage levels, with a median close to 5. Guest houses have a wider IQR, suggesting more variability in ICT adoption, with a median of around 5. Non-serviced accommodations show the most variability, with a median ICT usage level of approximately 4. This shows that the main trend of ICT use levels is similar across business types, but the spread is different, especially for non-serviced accommodations. This could mean these businesses have different ICT integration and dependence levels. This consistency in the distribution of ICT usage levels supports the evidence on ICT barriers (Table 3), emphasising the diverse ICT adoption patterns.

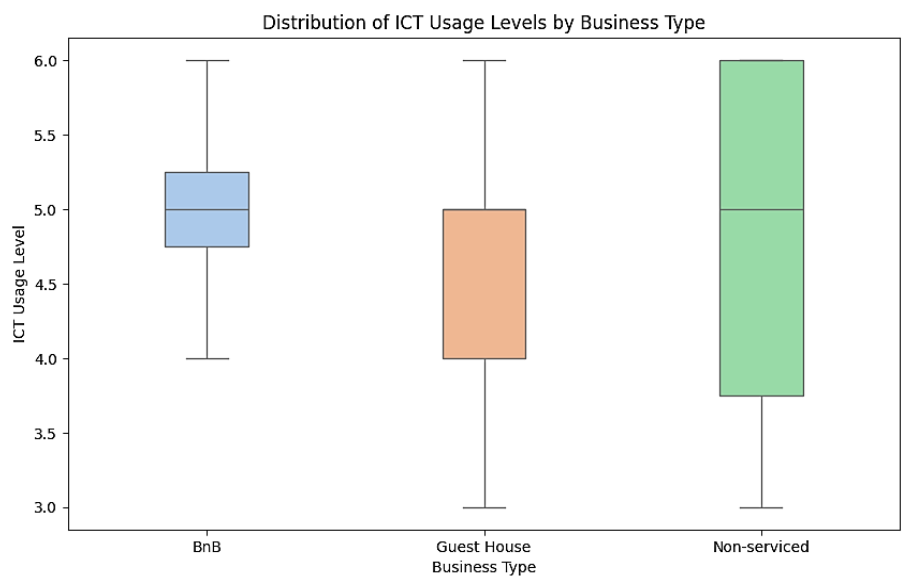

**Figure 2: Distribution of ICT Usage Levels Across the Business Types**

## 4.6 Discussion

The analysis of the data reveals several significant insights into the factors influencing ICT adoption in the small-scale hospitality sector. The critical role of education in driving ICT adoption stands out as a key finding. The higher levels of educational attainment among respondents correlate with greater computer literacy and a higher rating of ICT skills, which, in turn, enhance the overall ICT adoption rates. This underscores the importance of investing in education and targeted ICT training programs to bridge existing gaps and improve the digital readiness of businesses in this sector.

Another important finding is the proactive approach of businesses in leveraging ICT tools, particularly social media platforms like Facebook, for marketing purposes. This demonstrates a growing awareness of the benefits of digital marketing and the role of e-marketplaces in reaching a broader customer base. Educational programs focusing on digital marketing strategies could further empower these businesses to maximise their online presence and customer engagement.

The respondents' awareness of ICT benefits on various indicators such as years of ICT usage, impact on front office operations, and utilisation of ICT tools revealed that most respondents have been using ICT for 1-2 years (48%), with a notable positive impact on their business operations. All respondents acknowledged the beneficial changes brought about by ICT in their front office operations. ICT tools like Facebook, websites, and email for marketing and customer interaction were prevalent, indicating a high level of awareness and integration of ICT benefits in business practices. Access to ICT resources based on the availability of devices, internet access, and information storage methods show that most respondents (95%) had internet access and used various devices, such as personal computers, laptops, and smartphones, for business purposes. A significant portion of respondents offered internet access to their customers, either through pay-as-you-go or free Wi-Fi services. There is also a preference for storing and recording information electronically or using both electronic and traditional methods, underscoring the importance of ICT resources in daily business operations.

The Kruskal-Wallis H test on ICT usage levels categorised by years of ICT usage and by business types (BnBs, guest houses, and non-serviced accommodations indicated no significant differences in ICT usage levels for both groups. Specifically, the analysis based on years of ICT usage suggested that the duration of ICT usage does not significantly impact the level of ICT adoption. Similarly, the analysis based on business types showed no significant differences in ICT usage levels among the different types of accommodations. These findings suggest that other factors, such as access to resources, training, and possibly other variables, might play a more crucial role in influencing ICT adoption among the small-scale players in the area.

Establishing policies that uphold educational programs designed to enhance computer literacy and ICT skills for small-scale business owners and employees in the hospitality sector is imperative. Collaboration among government agencies, educational institutions, and industry stakeholders is crucial in developing and implementing ICT training programs customised for the industry's specific requirements. Encouraging continuous professional development in ICT will also ensure that business owners and employees stay abreast of the rapidly evolving technological landscape. This may entail providing regular training sessions, offering access to online learning platforms, and integrating ICT education into any existing business management programs.

## 5. Conclusion

The study provided valuable insights into the level of ICT literacy and adoption among the small-scale players in the hospitality industry in Mnquma Municipality. The findings revealed a high level of awareness and a positive perception of ICT benefits, with most businesses recognising the transformative potential of these technologies. However, the extent of ICT adoption varies, influenced by factors such as years of usage, access to resources, workforce educational level, and training availability. The role of education emerged as a key determinant in ICT adoption, with higher levels of educational attainment correlating with greater computer literacy and ICT skills, thereby facilitating better integration of digital tools into business operations. Despite the generally positive outlook, a significant portion of businesses remain either slightly dependent or not dependent on ICT, underscoring the need for targeted educational interventions to enhance ICT integration. While this study offers a robust understanding of ICT adoption in the hospitality sector sampled, it is not without limitations. The sample size and geographical constraints may limit the generalizability of the findings. Future research should aim to include a larger and more diverse sample to validate these results further. Emerging trends like artificial intelligence and blockchain could also be explored to ensure these businesses remain competitive in an increasingly digital landscape.